\documentclass[a4paper,10pt]{article}
\usepackage[cp1251]{inputenc}    
\usepackage[T2A]{fontenc}       
\usepackage[unicode]{hyperref} 

\usepackage{amssymb}
\usepackage{graphicx}
\usepackage{times}

\newcommand{\prob}{\mathrm{Prob}}
\newcommand{\eproof}{$\Box$}

\newtheorem{theorem}{Theorem}
\newtheorem{definition}{Definition}
\newtheorem{example}{Example}
\newtheorem{lemma}{Lemma}

\begin{document}

\title{Coding in the fork network in the framework of Kolmogorov complexity\thanks{This article is mostly a translation of the paper 
\emph{A.\,Romashchenko, A complexity version of the network coding problem.
Information Processes (electronic journal) 5 (2005) No. 1, pp. 20--28. }(in Russian).}}

\author{Andrei Romashchenko}

\maketitle

\begin{abstract}
Many statements of the classic information theory (the theory of Shannon's entropy) have natural counterparts in the algorithmic information theory (in the framework of Kolmogorov complexity). In this paper we discuss 
one simple instance of the parallelism between Shannon's and Kolmogorov's theories:
we  prove in the setting of Kolmogorov complexity an algorithmic  version of Wolf's characterization of admissible   rates for the fork network.
\end{abstract}

\section{Introduction: the Slepian--Wolf coding scheme}

Many  remarkable  similarities  between the probabilistic  and algorithmic  information theories were studied since the seminal paper of Kolmogorov \cite{Ko65}. 
In the present article we  discuss one particular example of the parallelism between Shannon's and Kolmogorov's frameworks for information theory. 
We study coding schemes for simple  multi-source networks (for so-called \emph{fork networks}) and show that Wolf's theorem from the classic information theory can be naturally translated in the framework of Kolmogorov complexity.

First of all, we remind the classic  Slepian--Wolf theorem and its algorithmic counterpart.   A special case of this theorem  (its non symmetric version) shows how an auxiliary source $\beta$ can be used for efficient compression of a source $\alpha$:
 \begin{theorem}[\cite{SW73}]\label{theorem-ws}
Let $(\alpha^i,\beta^i)$, $i=1,2,\ldots$ be a sequence of i.i.d. pairs of random variables jointly distributed
on some finite range $A\times B$. Then for every $\varepsilon>0$
there exist mappings 
 $$
 \begin{array}{rcccl}
 f^n & : & A^n & \to &\{0,1\}^{l(n)},\\
 g^n & : & \{0,1\}^{l(n)} \times B^n & \to & A^n
 \end{array}
 $$
such that
  $$\prob[(\alpha^1,\ldots,\alpha^n) = 
    g^n(f^n(\alpha^1,\ldots,\alpha^n),(\beta^1,\ldots,\beta^n))] 
     > 1-\varepsilon,$$ 
and $\lim\limits_{n\to\infty} \frac{l(n)}{H(\alpha|\beta)} = 1$.     
 \end{theorem}
Theorem~\ref{theorem-ws} has a clear intuitive meaning. The sender encodes  $n$  randomly chosen values of $\alpha^i$ in the most economic way.
The encoding function is denoted $f^n : A^n \to \{0,1\}^{l(n)}$. The receiver has to reconstruct the values of $\alpha^i$ given some additional information (the values of $\beta^i$ correlated with $\alpha^i$). The decoding function is denoted $g^n$.  The error probability must be bounded by some $\varepsilon>0$.
The aim is to minimize the length $l(n)$ of the transmitted  message.

Shannon's coding theorem claims that we can achieve the  error probability $\varepsilon$ with $l(n) = H(\alpha^1\ldots \alpha^n)+o(n)$, even if $\beta^i$ are not used in the decoding.  Slepian and Wolf show that given $\beta^i$ we can reduce the length of the message to 
$$
H(\alpha^1\ldots \alpha^n|\beta^1\ldots \beta^n)+o(n).
$$
What makes this theorem notrivial is that the values of $\beta^i$ are available only to the receiver and not the the sender,
i.e., $(\beta^1,\ldots,\beta^n)$ is an argument of  $g^n$ but not of $f^n$.

In the framework of Kolmogorov complexity a counterpart of Theorem~\ref{theorem-ws}   was proven by An.\,A.\,Muchnik, see \cite{Mu00,Mu02}:

 \begin{theorem}\label{theorem-mu}
For all strings   $a,b$ there exists a string $a'$ such that
  \begin{enumerate}
   \item $K(a'|a) = O(\log n)$,
   \item $K(a|a',b) = O(\log n)$,   
   \item $|a'| = K(a|b)$,
  \end{enumerate}
where $n=K(a)+K(b)$.
 \end{theorem}
  (See also a similar result  \cite[Theorem~3.11]{gacs};  an analogous technique was used in \cite{fl,bfl}.)
 In this theorem the string $a'$ plays the role of a  {\it code} that allows to ``easily'' reconstruct 
 $a$ given $b$. Moreover, the code $a'$ can be ``easily'' computed from $a$. 
 As usual in the theory of Kolmogorov complexity, the words ``easily computed'' mean that the corresponding
 conditional Kolmogorov complexity is bounded by $O(\log n)$. 

Loosely speaking, Theorem~\ref{theorem-mu} claims that among all \emph{almost shortest} programs that translate $b$ to $a$
there is one whose complexity conditional on $a$ is negligibly small.

Theorem~\ref{theorem-ws} is optimal  in the sense that the ratio
$\frac{l(n)}{H(\alpha|\beta)}$ cannot be made less than $1$.
Similarly,  in Theorem~\ref{theorem-mu} under conditions
(1) and (2) we have $|a'| \ge K(a|b) - O(\log n)$. 

The proofs of both Theorem~\ref{theorem-ws} and Theorem~\ref{theorem-mu}
consist in constructing suitable ``hash functions''; given the first source of information
we compute its fingerprint (a hash value), and then recover the initial value given this fingerprint and another
(auxiliary) source of information. However, the technical implementations of this idea in the proofs of 
Theorem~\ref{theorem-ws} and Theorem~\ref{theorem-mu} are pretty different.

\section{Fork networks}

Theorem~\ref{theorem-ws}  can be generalized for a larger class of communication networks.
Let us define the admissible rates for the ``\emph{fork networks}''.
 \begin{definition}\label{def-1}
Let a $k$-dimensional random variable $(\alpha_1,\ldots,\alpha_k)$ be distributed on a finite set  $A_1\times\ldots\times A_k$.
Denote by $(\alpha^i_1,\ldots,\alpha^i_k)$
($i=1,2,\ldots$) a sequence of  i.i.d. $k$-dimensional random variables, and let each of them be distributed as 
$(\alpha_1,\ldots,\alpha_k)$. 
A tuple of  $k$ reals 
$(r_1,\ldots,r_k)$ is called \emph{$\varepsilon$-admissible}  for the fork network with sources 
$\alpha_1,\ldots,\alpha_k$,
if for every $\delta>0$ and large enough  $n$ there exist functions  $f_1^n,\ldots,f_k^n$, $g^n$, 
  $$
 \begin{array}{rcccl}
  f_j^n &:& (A_j)^n &\to &\{0,1\}^{l_j(n)},\\
  g^n   &:& \{0,1\}^{l_1(n)+\ldots+l_k(n)} & \to & 
    A_1\times\ldots\times A_k
 \end{array}
 $$
such that  $l_j(n) \le (r_j+\delta)n$, and
   $$\prob[g^n(f_1^n(\bar \alpha_1),
             \ldots,
	      f_k^n(\bar \alpha_k)
	      ) = (\bar\alpha_1,\ldots,\bar\alpha_k)
	  ]>1-\varepsilon,
    $$
where  $\bar\alpha_j$ denotes the $n$-tuple $(\alpha_j^1,\ldots,\alpha_j^n)$ for each $j$.
 \end{definition}
\begin{figure}[t]
  \centering\includegraphics[height=80mm,width=120mm,clip]{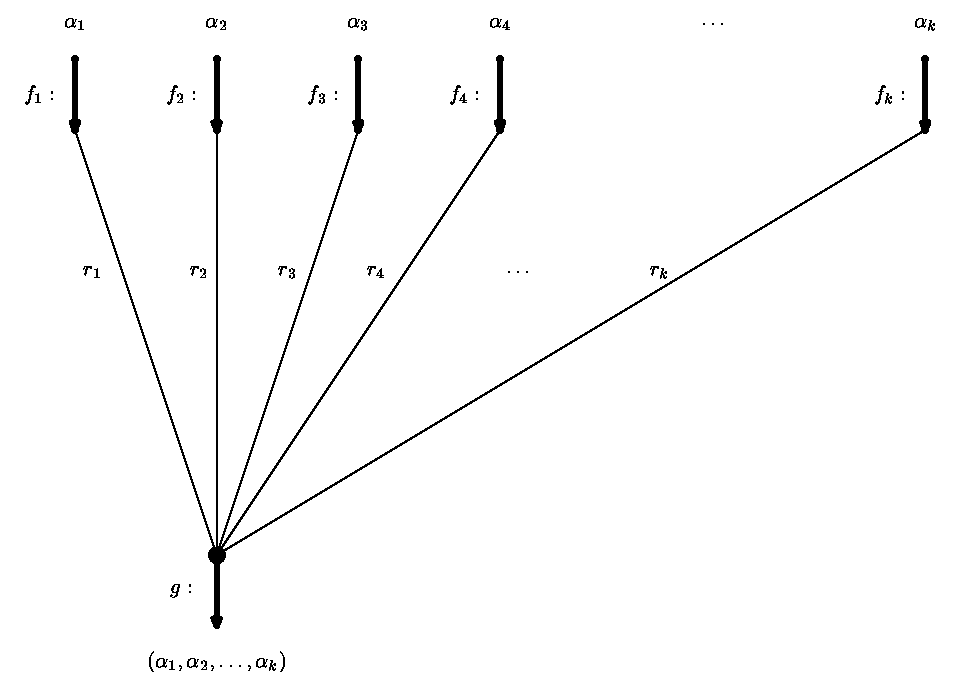}
  \begin{center}
    {\small Fig.~1. The fork network with $n$ sources}
  \end{center}
\end{figure}     
This definition corresponds to the information transmission in the network shown in Fig.~1. We are given $k$ correlated sources of information,
and their distribution is specified by the varables $\alpha_1,\ldots,\alpha_k$. The sources are  encoded independently by the block codes $f_j^n$.
The definition specifies the lengths of the encoded messages: the senders spend on average $r_j$ bits per each letter of the source $\alpha_j$. The receiver recovers the values of all $k$ sources with the decoding function $g^n$, and the probability of error
must be less that $\varepsilon$.

The set of all $\varepsilon$-admissible rates can be characterized in terms of entropies involving the random variables $\alpha_j$. 

\smallskip
\emph{Notation:}
{Let  $(\alpha_1,\ldots,\alpha_k)$ be a tuple of jointly distributed random variables.
In what follows we denote by $\alpha_W$  the tuple of random variables $\alpha_j$ for all $j\in W$, and by
$\alpha_{\neg W}$ the tuple of random variables $\alpha_j$ for $j\not\in W$.
For example, if $k=5$ and $W=\{1,2,5\}$, then $\alpha_W$ denotes $(\alpha_1,\alpha_2,\alpha_5)$
and $\alpha_{\neg W}$ denotes $(\alpha_3,\alpha_4)$. For $W=\emptyset$
we suppose that $\alpha_W$ is a constant (a random variable with zero entropy).
In particular, if $W=\{1,\ldots,k\}$  and $\neg W=\emptyset$, then
$H(\alpha_W|\alpha_{\neg W})=H(\alpha_W)$.
}
\smallskip

Now we can formulate Wolf's theorem that characterizes the set of admissible rates for the fork networks.
 \begin{theorem}[J.K. Wolf \cite{Wo74}; see also \cite{CK85}]\label{theorem-fork-sh}
 
For every $k$-tuple of  jointly distributed random variables $(\alpha_1,\ldots,\alpha_k)$
and for every  $\varepsilon>0$,

(i) \textup[the necessary condition\textup]  for every $\varepsilon$-admissible  tuple of reals $(r_1,\ldots,r_k)$
and for all $W\subset \{1,\ldots,k\}$ it holds
 $$\sum\limits_{j\in W} r_j \ge H(\alpha_W|\alpha_{\neg W}),$$

(ii) \textup[the sufficient condition\textup]  
{if} for each set $W\subset \{1,\ldots,k\}$ it holds
 $$\sum\limits_{j\in W} r_j > H(\alpha_W|\alpha_{\neg W}),$$
then the tuple of reals $(r_1,\ldots,r_k)$ is $\varepsilon$-admissible for the fork network.
 \end{theorem}

 \begin{example}
For $k=2$ Theorem~\ref{theorem-fork-sh} implies that 
a pair $(r_1,r_2)$ is  $\varepsilon$-admissible  for the network with sources 
$(\alpha_1,\alpha_2)$, only if 
  $$
  \begin{array}{rcl}
   r_1 &\ge& H(\alpha_1|\alpha_2),\\
   r_2 &\ge& H(\alpha_2|\alpha_1),\\
   r_1+ r_2 &\ge& H(\alpha_1,\alpha_2).
   \end{array}
  $$
 Similarly, the conditions
  $$
  \begin{array}{rcl}
   r_1 &>& H(\alpha_1|\alpha_2),\\
   r_2 &>& H(\alpha_2|\alpha_1),\\
   r_1+ r_2 &>& H(\alpha_1,\alpha_2).
   \end{array}
  $$
are enough to guarantee that a pair $(r_1,r_2)$ is  $\varepsilon$-admissible. This special case of Theorem~\ref{theorem-fork-sh} is
 the general statement of the Slepian--Wolf  theorem, \cite{SW73}. 
  If $r_2=\infty$, then the only remaining condition is $r_1\ge H(\alpha_1\ | \alpha_2)$, so we get the statement of 
  Theorem~\ref{theorem-ws} as a special case.
 
 \end{example} 

In what follows we  prove a counterpart of Theorem~\ref{theorem-fork-sh} for Kol\-mo\-go\-rov complexity. Technically,
we give a criterion for the following property of a tuple of binary strings $a_1,\ldots,a_n$ (which is a counterpart of the \emph{admissibility}
property from Definition~\ref{def-1} adapted to the Kolmogorov's theory):
$$
\begin{array}{l}
\mbox{{\it For a tuple of strings }}a_1,\ldots,a_k 
\mbox{ {\it and a tuple of integers }}r_1,\ldots,r_k \\
\mbox{{\it there exist strings }} a'_1,\ldots, a'_k, 
\mbox{{\it such that }}\\
\ \ \ (1) \ |a'_j|\le r_j,\ j=1,\ldots,k,\\
\ \ \ (2) \ K(a'_j|a_j)\le C_0\log n,\ j=1,\ldots,k, \\
\ \ \ (3) \  K(a_1,\ldots,a_k|a'_1,\ldots,a'_k)<C_0\log n,\\
{\it where }\ n=|a_1|+\ldots+|a_k|.
\end{array}
\eqno(*)
$$

\begin{theorem}[main result]\label{theorem-main}
\rule{0pt}{10pt}

\noindent
\textup{[The necessary condition]:}
For all integer $k>0$ and $C_0>0$ there exists a constant
$C_1$ such that for all strings $a_1,\ldots,a_k$ (with total length $n=|a_1|+\ldots+|a_k|$) and all integers $r_1,\ldots,r_k$,
property (*) holds \emph{only if} for every non empty set
$W\subset \{1,\ldots,k\}$ 
 $$\sum\limits_{j\in W} r_j \ge K(a_W|a_{\neg W})-C_1\log n.$$

\noindent
\textup{[The sufficient condition]:}
For all integer $k>0$ and $C_2>0$  there exists a constant
$C_0$ such that for all strings $a_1,\ldots,a_k$ and integers $r_1,\ldots,r_k$
property (*) holds if for every nonempty set  
$W\subset \{1,\ldots,k\}$ 
 $$\sum\limits_{j\in W} r_j \ge K(a_W|a_{\neg W})+C_2\log n.$$

\end{theorem}
\emph{Notation:}
In this theorem we use the notation $a_W$, which stands for a tuple of all strings $a_j$ for $j\in W$. Similarly,
$a_{\neg W}$ stands for a tuple of all strings $a_j$ for $j\not\in W$.
For the empty  $W$ we denote by $a_W$ the empty word. 
In particular, if $W=\{1,\ldots,k\}$ and $\neg W=\emptyset$, then
$K(a_W|a_{\neg W}) = K(a_W) + O(1)$.
\begin{example}
For $k=2$ this theorem gives the necessary and sufficient conditions 
  $$
  \begin{array}{rcl}
   r_1 &\ge& K(a_1|a_2)+O(\log n),\\
   r_2 &\ge& K(a_2|a_1)+O(\log n),\\
   r_1+ r_2 &\ge& K(a_1,a_2)+O(\log n).
   \end{array}
  $$
\end{example}  
For $k=2,3$ Theorem~\ref{theorem-main} was proven in \cite{Iz04}. 
In the present paper we prove this theorem for all integer $k>0$.

The standard proof of Theorem~\ref{theorem-fork-sh} (see \cite{CK85})
cannot be translated in the language of Kolmogorov complexity.  The crucial point is that 
the proof in \cite{CK85} employs the principle of \emph{time sharing}, 
which does not apply 
in the framework of Kolmogorov complexity. 
We prove Theorem~\ref{theorem-main} using the following version of Muchnik's theorem 
(which is somewhat stronger than Theorem~\ref{theorem-mu}):

 \begin{theorem}[\cite{Mu02}]\label{theorem-mu-xt}
For every integer $k>0$ there exists a number  $C=C(k)$ with the following property.
Let $x_0,x_1,\ldots,x_k$ be binary strings, $n=|x_1|+\ldots+|x_k|$,
and $r$ be a number less than  $|x_0|$. Then there exists a string $y$ such that
 \begin{enumerate}
 \item[(1)] $|y|=r$, 
 \item[(2)] $K(y|x_0) \le C\log n$,
 \end{enumerate}
and 
 $K(y|x_j)\ge \min\{ K(x_0|x_j), r\} -C\log n$
 for every $j=1,\ldots,k$.
 \end{theorem}
Informally, Theorem~\ref{theorem-mu-xt} claims that we can extract (with only a logarithmic advice) from 
a string  $x_0$  a fingerprint of length $r$ that looks maximally ``random'' given each of  the strings $x_j$ as a condition.

\smallskip
\noindent
\textbf{Remark 1:}
Since $K(y|x_0) = O(\log n)$,  for every $j$  we have
 $$
 K(y| x_j) \le K(x_0|x_j) + O(\log n).
 $$

\smallskip
\noindent
\textbf{Remark 2:} If $r> K(x_0|x_j)$, then $x_0$ can be completely reconstructed given $y$ and $x_j$ (and some logarithmic advice), i.e., 
 $$
 K(x_0| y, x_j) = O(\log n).
 $$
 Indeed, in this case we have $K(y|x_j) = K(x_0|x_j)+O(\log n)$, so
  $$
  \begin{array}{rcl}
    K(x_0| y, x_j)&=& K(x_0, y|x_j) - K(y| x_j) + O(\log n)\\
      			&=& K(x_0 | x_j) - K(y| x_j) + O(\log n)\\
      			&=&  O(\log n).
  \end{array}
  $$

 \smallskip
\noindent
\textbf{Remark 3:}
{Theorem~\ref{theorem-mu-xt}  implies   Theorem~\ref{theorem-mu}.}
Indeed, let us  apply Theorem~\ref{theorem-mu-xt} for $k=1$, with $x_0=a$,
$x_1=b$, $r_1=K(a|b)$. We obtain a string $y$ such that
\begin{enumerate}
 \item $|y|=r_1$,
 \item $K(y|a)=O(\log n)$,
 \item $K(y|b)\ge K(a|b)-O(\log n)$.
 \end{enumerate}
These conditions imply that  $K(a|y,b)=O(\log n)$.  
Thus,  we may let $a'=y$. \eproof

\section{Proof of theorem~\ref{theorem-main}}

For the sake of brevity, we use the following asymptotic notation: 
 $$
\begin{array}{rcl}
  F(n)\le_n G(n) & \leftrightharpoons & F(n)\le G(n) + O(\log n),\\
  F(n)\ge_n G(n) & \leftrightharpoons & G(n)\le F(n) + O(\log n),\\  
  F(n) =_n G(n) & \leftrightharpoons & F(n)= G(n) + O(\log n).  
\end{array}
 $$

{\bf Proof of the necessity condition}: Let  $W\subset\{1,\ldots,k\}$ be any set of indices, and  $\neg W = \{1,\ldots,k\}\setminus W$.
By the condition of the theorem, all strings $a_i$ have logarithmic complexity conditional on 
the tuple $\langle a'_1,\ldots,a'_k\rangle$. It follows that 
$K(a_W | a'_W,a_{\neg W})=_n 0$. Hence,  
complexity of the tuple $a'_W$ cannot be less than the conditional complexity 
$K(a_W |a_{\neg W})$.  On the other hand,  Kolmogorov complexity of  $a'_W$ is not greater than  $r_i$ for each $i\in W$, and we are done.
More formally this argument can be presented as a chain of inequalities:
 $$
 \begin{array}{rcl}
  K(a_W|a_{\neg W}) & \le_n & K(a'_W) + K(a_W|a'_W,a_{\neg W}) \le_n \\
                     &\le_n&       K(a'_W) + K(a_W|a'_W,a'_{\neg W}) \le_n \\
                     &\le_n&       K(a'_W) \le_n \sum\limits_{j\in W}r_j.
  \end{array}				 
  $$

{\bf Proof of the sufficiency condition}: We prove the theorem by induction on $k$. 
To make the inductive step work, we need to reformulate the  theorem and make it somewhat stronger:

\medskip
\noindent
{\bf Inductive claim:}
{\it
For every integer $k>0$ and for all $C_2>0$ there exists a number $C_0$ with the following property.
Let $a_1,\ldots,a_k,b$ be binary strings and  $r_1,\ldots,r_k$ be integers,
denote $n=|a_1|+\ldots+|a_k|+|b|$.
Assume that for every nonempty $W\subset \{1,\ldots,k\}$ it holds $\sum\limits_{j\in W} r_j \ge K(a_W|a_{\neg W},b)+C_2\log n$. Then it follows
that there exist binary strings $a'_1,\ldots, a'_k$ such that
  \begin{enumerate}
  \item $|a'_j|\le r_j$, $j=1,\ldots,k$,
  \item $K(a'_j|a_j)\le C_0\log n$, $j=1,\ldots,k$,
  \item $K(a_1,\ldots,a_k|a'_1,\ldots,a'_k,b)<C_0\log n$.
  \end{enumerate} 
}

\medskip

\noindent
The difference between this claim and  (*) is  a new parameter $b$. 

For $k=1$ the inductive claim follows immediately from Theorem~\ref{theorem-mu}.
Let us perform the inductive step. Fix some binary strings $a_1,\ldots,a_k,a_{k+1},b$.
From Theorem~\ref{theorem-mu-xt} it follows that there exists a string $a'_{k+1}$
such that
 \begin{itemize}
 \item $|a'_{k+1}|\le r_{k+1}$, 
 \item $K(a'_{k+1}|a_{k+1}) \le C\log n$,
 \end{itemize}
and for every nonempty $W\subset \{1,\ldots,k\}$
 $$K(a'_{k+1}|a_W,b)\ge \min\{ K(a_{k+1}|a_W,b), r_{k+1}\} -C\log n \leqno(**)$$
(the value of $C$ depends on $k$ but not on $a_i$ and $b$).

We are going to use  the inductive hypothesis with the tuple of strings 
$$a_1,\ldots,a_k, b':=\langle a'_{k+1},b\rangle$$ 
and the tuple of integers 
$r_1,\ldots r_k$. 
To this end we should verify that the \emph{inductive claim} is applicable to these strings, 
i.e., we need to prove the following lemma.
 \begin{lemma}\label{lemma}
 There exists a  $C'_2=C'_2(k,C_2)$ such that for every non-empty
$V\subset \{1,\ldots,k\}$ and  its complement $\neg V= \{1,\ldots,k\}\setminus V$ it holds
 $$\sum\limits_{j\in V} r_j \ge K(a_V|a_{\neg V},b')+C'_2\log n.$$
 \end{lemma}
{\bf Proof:}
We consider separately two cases. 

\emph{Case 1:} Assume that $r_{k+1}\ge K(a_{k+1}|a_{\neg V},b)$. From Remark~2 we know that 
 $$
 K(a_{k+1}|a_{k+1}',a_{\neg V},b) =_n 0.
 $$
Hence, 
$$
 K(a_V|a_{\neg V},a_{k+1}', b) \le_n K(a_V|a_{\neg V},a_{k+1}, b) \le_n \sum\limits_{j\in V} r_j 
$$
(the last inequality is a part of the condition of the inductive claim). 

\smallskip

\emph{Case 2:} Now we assume that $r_{k+1}< K(a_{k+1}|a_{\neg V},b)$. 
We are given the condition
 $$r_{k+1}+\sum\limits_{j\in V} r_j  \ge_n K(a_{k+1},a_V|a_{\neg V},b).$$
Using the Kolmogorov--Levin theorem, we can reformulate this inequality as
 $$
 \begin{array}{rcl}
 r_{k+1}+\sum\limits_{j\in V} r_j  &\ge_n&   K(a_{k+1}'|a_{\neg V},b)
    + K(a_{k+1}|a_{k+1}',a_{\neg V},b)\\
   && +K(a_V, |a_{k+1},a_{\neg V},b).  
   \end{array}\leqno(\mbox{***})
 $$
Then we get from (**) 
 $$K(a_{k+1}'|a_{\neg V},b)=_n r_{k+1}.$$
So (***) rewrites to
 $$
 r_{k+1}+\sum\limits_{j\in V} r_j  \ge_n   r_{k+1}+ K(a_{k+1}|a_{k+1}',a_{\neg V},b) +K(a_V, |a_{k+1},a_{\neg V},b),
 $$
 which implies 
  $$
\sum\limits_{j\in V} r_j  \ge_n  K(a_V, |a_{k+1},a_{\neg V},b),
 $$
 and we are done.
\eproof

With Lemma~\ref{lemma} we can apply the inductive hypothesis. We obtain some strings 
 $a'_1,\ldots,a'_k$ such that
  \begin{enumerate}
  \item $|a'_j|\le r_j$, $j=1,\ldots,k$,
  \item $K(a'_j|a_j)\le C'_0\log n $, $j=1,\ldots,k$,
  \item $K(a_1,\ldots,a_k|a'_1,\ldots,a'_k,a'_{k+1},b)\le C'_0\log n$
  \end{enumerate} 
for some $C'_0=C'_0(k,C_2)$.
It remains to show that
 $$K(a_{k+1}|a'_1,\ldots,a'_k,a'_{k+1},b)\le C_0 \log n$$
for some $C_0\ge C'_0$ (which may depend on $k$ and $C_2$). 
To this end, it is enough to prove
$K(a_{k+1}|a_1,\ldots,a_k,a'_{k+1},b)=_n 0$.
For the sake of brevity, we use the asymptotic notation: 
$$
\begin{array}{l}
     K(a_{k+1}|a_1,\ldots,a_k,a'_{k+1},b) =_n\\
     =_n
     K(a_{k+1},a'_{k+1}|a_1,\ldots,a_k,b)
   - K(a'_{k+1}|a_1,\ldots,a_k,b) \\
   =_n K(a_{k+1}|a_1,\ldots,a_k,b)
   - K(a'_{k+1}|a_1,\ldots,a_k,b) \\
   =_n K(a_{k+1}|a_1,\ldots,a_k,b) - \min\{K(a_{k+1}|a_1,\ldots,a_k,b), r_{k+1}\}
    =_n 0.
   \end{array}
 $$
\eproof

\section{Conclusion}

It seems natural to ask whether 
a version of Theorem~\ref{theorem-main} holds for resource bounded versions of Kolmgorov complexity, e.g., for programs  running in polynomial time or polynomial space.
Recently M.~Zimand proved a  variant of Thereom~\ref{theorem-main} where the encoding procedures $a'_j = Enc_j(a_j)$, $j=1,\ldots,k$ can be performed by  probabilistic polynomial time algorithms, see \cite{Zimand}. It seems unlikely that   the optimal  lengths of ``codewords'' $a'_j$ and polynomial time encoding could be combined also with polynomial time  decoding $(a'_1\ldots a'_k)\mapsto (a_1\ldots a_k)$. 


\bigskip

\noindent
\textbf{Acknowledgments.} The author is grateful to Marius Zimand, who pointed out an error in the first version of this paper.


\begin{thebibliography}{11}

\bibitem{Ko65} Kolmogorov A.N.,
  \emph{Three approaches to the quantitative definition of information},
  {Problems of information transmission}, 1(1), 1--7, 1965.





\bibitem{SW73} Slepian~D., Wolf~J.K., Noiseless coding of correlated
information sources. {\it IEEE Transactions on Information Theory}, 
19, 471--480, 1973.

\bibitem{Wo74} Wolf~J.K., Data reduction for multiple correlated
sources. In: {\it Proc. of the Fifth Colloquium on Microwave Communication.}
Budapest, 287--295, 1974.









\bibitem{gacs} Bennett\,C.H., G\'acs\,P., Li\,M., Vit\'anyi\,P. M., Zurek\,W.\,H. \emph{Information distance}. 
IEEE Transactions on Information Theory, 44(4), 1998,1407--1423.


\bibitem{fl} Fortnow L., Laplante S. 
\emph{Nearly optimal language compression using extractors}. In Proc. STACS, 1998. 84--93.


\bibitem{Mu00} Muchnik~An.A., Semenov~A.L. 
\emph{Multi-conditional Descriptions and codes in Kolmogorov complexity.}
{Electronic Collocuium on Computational Complexity (ECCC)}, 7(15), 2000.

\bibitem{bfl}Buhrman H., Fortnow L., Laplante S. 
\emph{Resource-bounded Kolmogorov complexity revisited}. SIAM Journal on Computing.  31(3), 887--905, 2001.


\bibitem{Mu02} Muchnik, A.A. \emph{Conditional complexity and codes.} Theoretical Computer Science, 271(1), 97--109, 2002.




\bibitem{Iz04} Izmailova A.A., Information transmission in the fork network with bounded channel capacities.
Master thesis. Moscow State University, 2004. In Russian.
({\small Измайлова~А.А.,  Передача сообщений в вилочной сети
с ограниченными пропускными способностями каналов. {\it Дипломная работа.}
Москва, МГУ им.~Ломоносова. 2004.})



\bibitem{CK85}  Csiszar I., K{\"o}rner J., 
\emph{Information theory: coding theorems for discrete memoryless systems}, Cambridge University Press,
{2011}.


\bibitem{Zimand} Zimand, M.  \emph{Kolmogorov complexity version of Slepian-Wolf coding.}  arXiv:1511.03602 (2015).

\end{thebibliography}
\end{document}